\documentclass[twocolumn,showpacs,prb]{revtex4}
\usepackage{graphicx}
\usepackage{dcolumn}
\usepackage{bm}
\begin{document}

\title{Thermodynamic Evidence for Water as a Quantum Mechanical Liquid}
\author{A. Widom and J. Swain}
\affiliation{Physics Department, Northeastern University, Boston MA USA}
\author{Y.N. Srivastava}
\affiliation{Physics Department, University of Perugia, Perugia IT}
\author{S. Sivasubramanian}
\affiliation{Nanoscale Science \& Engineering Center for High-rate 
Nanomanufacturing\\ Northeastern University, Boston MA USA.}
\author{D. Drosdoff}
\affiliation{Department of Mathematics and Physics, 
North Carolina Central University, Durham NC USA}

\begin{abstract}

We consider general theoretical models of water and in particular the
nature of the motions of the hydrogen nuclei. If the motion of hydrogen nuclei
is classical, then the thermodynamic pressure equation of state for heavy
water wherein the hydrogen nuclei are deuterons is identical to the
pressure equation of state for light water wherein the hydrogen nuclei are protons.
Since the experimental thermodynamic phase diagram for light water is
clearly measurably different from the experimental thermodynamic phase diagram
for heavy water, one may deduce that the motions of hydrogen nuclei are
quantum mechanical in nature. This conclusion is in physical agreement with a
recent analysis of X-ray, neutron and deep inelastic neutron scattering data.

\end{abstract}


\maketitle

\section{Introduction \label{intro}}

The study of the applications of quantum mechanics to water is of long
standing interest. Recent important
studies\cite{Reiter:2004, Pantalei:2008, Flammini:2009, Pietropaolo:2009}
of the momentum distributions and position correlations\cite{Soper:2008}
in light water \begin{math} {\rm H_2O} \end{math} and in heavy water
\begin{math} {\rm D_2O} \end{math} have provided strong evidence that the
proton motions in light water and deuteron motions in heavy water are
{\em in reality quantum mechanical}. The experimental data arise from 
X-ray and neutron scattering experiments. In particular, from deep inelastic 
neutron scattering from protons, one may deduce the momentum distribution 
of protons within water and it is found to differ substantially from the 
Maxwellian distribution expected from classical proton motions. In fact, quantum 
fluctuations in the proton velocity dominate the sometimes presumed classical thermal 
fluctuations.

Our purpose here is to provide experimental {\it thermodynamic evidence} that 
the motions of hydrogen atoms in liquid water {\em cannot be adequately described} 
by classical mechanics. The thermodynamic data are hardly of recent vintage.
However, the thermal data have not been previously analyzed in detail for the 
purpose of assessing the contribution of quantum mechanics to the thermodynamic 
properties of water. As a theoretical model, we at first presume a Hamiltonian which consists 
of kinetic energy plus Coulomb potential energy; It is  
\begin{eqnarray}
{\cal H}&=&{\cal K}+{\cal V},
\nonumber \\ 
{\cal K}&=&\left\{\sum_a \frac{P_a^2}{2M_a}+\sum_j \frac{p_j^2}{2m}\right\}, 
\nonumber \\ 
{\cal V}&=&e^2\left\{\sum_{a<b} \frac{z_az_b}{R_{ab}}+
\sum_{i<j}\frac{1}{r_{ij}}-\sum_{aj}\frac{z_a}{|{\bf r}_j-{\bf R}_a|}\right\},
\label{intro1}
\end{eqnarray} 
wherein the indices \begin{math} a,b,\ldots \end{math} refer to the nuclei 
of hydrogen and oxygen and the indices \begin{math} i,j,\ldots \end{math} refer to 
the electrons. We also presume that the hydrogen nuclei obey classical mechanics to 
a sufficient degree of accuracy; i.e.  
\begin{equation} 
\left<\frac{P^2}{2M}\right>=\frac{3}{2}k_BT 
\ \ \ ({\rm Classical\ Hydrogen\ Nucleus}).
\label{intro2}
\end{equation}
In Sec.\ref{teos} we prove the following:
\par \noindent 
{\bf Theorem:} Under the presumptions of classical motions for the hydrogen 
nuclei and with the free energy per molecule \begin{math} f(v,T) \end{math} 
obeying  
\begin{equation} 
df=-Pdv-sdT,
\label{intro3}
\end{equation}
the pressure equation of state \begin{math} P(v,T) \end{math} will be 
{\em identical} for {\em light water} with proton hydrogen nuclei and
{\em heavy water} with deuteron hydrogen nuclei. Here,  
\begin{math} v \end{math} represents the volume per molecule.
As a consequence, light water and heavy water are predicted to have identical 
phase diagrams. These conclusions remain valid for more fully quantum electrodynamic 
microscopic models as shown in Sec.\ref{mdi}.

The experimental evidence\cite{Harvey:2002} is that the theorem fails in the 
laboratory. For example, at atmospheric pressure, the melting temperature of 
light ice into light water is 
\begin{math} T_{m1} \approx 0{\rm \ ^oC}  \end{math}. At atmospheric pressure 
the melting temperature of heavy ice into heavy water is 
\begin{math} T_{m2} \approx 4{\rm \ ^oC}  \end{math}. 
In the concluding Sec.\ref{conc}, the {\em modifications to the incorrect 
presumption of classical motion} required by thermodynamic experiments is 
discussed.

\section{Thermal Equations of State \label{teos}}

With \begin{math} N \end{math} molecules in a volume 
\begin{math} V \end{math}, the free energy of water is found from 
\begin{eqnarray}
F&=&-k_BT\ln Tr_{(V,N)}\left\{ e^{-{\cal H}/k_BT} \right\}, 
\nonumber \\ 
dF&=&-PdV+\mu dN-SdT.
\label{teos1}
\end{eqnarray}
In the thermodynamic limit, the free energy per molecule 
\begin{equation}
f(v,T)=\lim_{N\to \infty } \frac{F(V=Nv,N,T)}{N}
\label{teos2}
\end{equation}
obeys Eq.(\ref{intro3}). 

Consider the kinetic energy of the hydrogen nuclei 
\begin{equation}
{\cal K}_{H}=\frac{1}{2M}\sum_{c=1}^{\cal N} P_c^2 
\ \ {\rm wherein}\ \ {\cal N}=2N.
\label{teos3}
\end{equation}
As an operator, one may consider the Hamiltonian in Eq.(\ref{intro1}) as 
a function of the hydrogen nuclear mass \begin{math} M \end{math}. Hence, the 
identity
\begin{equation}
{\cal K}_{H}=-M\left(\frac{\partial {\cal H}}{\partial M}\right) .
\label{teos4}
\end{equation}
By a general theorem of statistical mechanics\cite{Landau:1980}, 
one has 
\begin{equation}
\left<{\cal K}_{H}\right>=-M\left(\frac{\partial F}{\partial M}\right)_{V,N,T}
=\frac{3}{2}{\cal N}k_BT, 
\label{teos5}
\end{equation}
wherein Eq.(\ref{intro2}) has been invoked. Regarding the proton or the deuteron 
as the nucleus of the hydrogen atom, it is important to note that the proton is a 
Fermion and the deuteron is a Boson. If the nucleus undergoes classical motion, 
then the whole notion of quantum statistics plays no role in the thermal equations  
of state. The free energy difference between light water and heavy water then can 
be found by integrating Eq.(\ref{teos5}) from the proton mass to the deuteron mass
(twice the proton mass to the precision we need here\cite{codata}).
In detail, 
\begin{eqnarray}
\Delta F = 
\int_{M_1}^{M_2}\left(\frac{\partial F}{\partial M}\right)_{V,N,T}dM ,
\nonumber \\ 
\Delta F = -\frac{3}{2}{\cal N}k_BT\int_{M_1}^{M_2}\left(\frac{dM}{M}\right),
\nonumber \\ 
\Delta F = 
-\frac{3}{2}{\cal N}k_BT\ln\left(\frac{M_2}{M_1}\right),
\nonumber \\ 
\Delta F = -\frac{3}{2}{\cal N}k_BT\ln 2=
-(3\ln 2) Nk_BT.  
\label{teos6}
\end{eqnarray}
The difference in classical theory between light water and heavy water thereby 
amounts to 
\begin{equation}
\Delta f=-(3\ln 2) k_BT.
\label{teos9}
\end{equation}
If the hydrogen nucleus moves via classical mechanics, then the only 
thermodynamic difference between light and heavy water is the  
entropy change per molecule 
\begin{equation}
\Delta s=-\left(\frac{\partial \Delta f}{\partial T}\right)_v
=(3\ln 2)k_B.
\label{teos10}
\end{equation}
Eq.(\ref{teos10}) in no way changes the pressure equation of state 
\begin{math} P(v,T)  \end{math}, 
\begin{equation}
\Delta P=
-\left(\frac{\partial \Delta f}{\partial v}\right)_T=0, 
\label{teos11}
\end{equation}
so that the theorem in Sec.\ref{intro} holds true. On the other hand, 
the difference between light water and heavy water is measurable in the 
laboratory. The question then arises as to which presumptions of the 
theory are incorrect.

\section{More Detailed Interactions \label{mdi}}

In order to consider quantum electrodynamic interactions, one may include
the vector potential in the Coulomb gauge,
\begin{equation}
{\bf B}({\bf r})=curl{\bf A}({\bf r})\ \ {\rm and}\ \ div{\bf A}({\bf r})=0.
\label{mdi1}
\end{equation}
Eq.(\ref{intro1}) must be replaced by 
\begin{eqnarray}
{\cal H} = {\cal K}+{\cal V}+{\cal H}_{\rm rad},
\nonumber \\ 
{\cal K} = 
\sum_a \frac{\big({\bf P}_a-(z_ae/c){\bf A}({\bf R}_a)\big)^2}{2M_a} 
\ \ \ \ \ \ \ 
\nonumber \\ 
+\sum_j \frac{\big({\bf p}_j+(e/c){\bf A}({\bf r}_j)\big)^2}{2m}\ , 
\nonumber \\ 
{\cal V} = e^2\left\{\sum_{a<b} \frac{z_az_b}{R_{ab}}+
\sum_{i<j}\frac{1}{r_{ij}}-\sum_{aj}\frac{z_a}{|{\bf r}_j-{\bf R}_a|}\right\}\ ,
\nonumber \\ 
{\cal H}_{\rm rad} = \frac{1}{8\pi}
\int \left( {\bf E}({\bf r})^2+{\bf B}({\bf r})^2 \right)d^3{\bf r},
\label{mdi2}
\end{eqnarray} 
wherein the transverse field equal time commutation relations read 
\begin{eqnarray}
\frac{i}{\hbar c}\left[A_k({\bf r}),E_l({\bf r}^\prime)\right] 
= 4\pi \Delta_{kl}({\bf r}-{\bf r}^\prime ), 
\nonumber \\ 
{\bf \Delta }({\bf r}) = 
\int \left({\bf 1}-\frac{\bf kk}{k^2}\right)
e^{i{\bf k\cdot r}}\left[\frac{d^3 {\bf k}}{(2\pi )^3}\right].
\label{mdi3}
\end{eqnarray}
The hydrogen nuclear velocity \begin{math} {\bf V} \end{math} is now given by  
\begin{equation}
M{\bf V}={\bf P}-\frac{e}{c}{\bf A}({\bf R}),
\label{mdi4}
\end{equation}
yielding a mean kinetic energy 
\begin{eqnarray}
\frac{1}{2}M\left<V^2\right>&\ge& \frac{3}{2}k_BT,
\nonumber \\ 
-M\left(\frac{\partial F}{\partial M}\right)_{V,N,T}
&\ge& \frac{3}{2}{\cal N}k_BT=3Nk_BT. 
\label{mdi5}
\end{eqnarray}
In Eq.(\ref{mdi5}) for the hydrogen nuclear kinetic energy, equality 
holds true when the hydrogen nuclear motions are classical and the 
inequality holds true when the hydrogen nuclear motions are quantum 
mechanical. The inequality in Eq.(\ref{mdi5}) is proved in 
Appendix \ref{ine}.  

Only in the classical case will heavy water have the same 
pressure equation of state \begin{math} P(v,T) \end{math} as light water. 
It is experimentally\cite{Harvey:2002} clearly the case that the phase 
diagram for heavy water differs from that of light water providing strong 
evidence for quantum mechanical hydrogen nuclear motions.

\section{Conclusion \label{conc}}

For quantum electrodynamic general theoretical models of water
we have shown that the notion of classically moving hydrogen nuclei
is in conflict with the experimental differences in the
thermodynamic phase diagrams of light and heavy water.
This conclusion is in physical agreement with recent results 
based on deep inelastic neutron 
scattering\cite{Reiter:2004, Pantalei:2008, Flammini:2009, Pietropaolo:2009} 
and on neutron and X-ray scattering\cite{Soper:2008}. 

The central point is that for a typical motion frequency 
\begin{math} \Omega  \end{math}, as given in Eq.(\ref{ine2_prime}) below,
a necessary condition for classical motion is that 
\begin{equation}
\beta \Omega \equiv \frac{\hbar \Omega }{k_BT}\ll 1 
\ \ \ ({\rm Classical\ Motion})
\label{conc1}
\end{equation}
which fails by a large margin for proton and/or deuterium motion 
in water. 

In more quantitative detail, suppose one defines a proton {\it quantum 
noise temperature } \begin{math} \tilde{T} \end{math} in terms of the 
mean proton kinetic energy 
\begin{equation}
\frac{1}{2}M\left<|{\bf V}|^2\right>\equiv \frac{3}{2}k_B\tilde{T}. 
\label{conc2}
\end{equation}
In general, \begin{math} \tilde{T}\ge T  \end{math} as proved in 
Appendix \ref{ine}. 
If the quantum noise temperature is equal (to a sufficient degree of accuracy) 
to the thermal temperature, \begin{math} \tilde{T}\approx T  \end{math}, then 
the proton motion may be presumed to be classical. The 
experimental number\cite{Pantalei:2008,Flammini:2009} for 
light water in the neighborhood  of room temperature is 
\begin{equation}
\tilde{T}\approx 3.7\ T
\ \ \ {\rm at}\ \ \ T\approx 300\ ^oK.
\label{conc3}
\end{equation}
Thus, the quantum noise in the proton velocity dominates 
the purely classical thermal fluctuations in the proton 
velocity for normal laboratory water samples. The large quantum mechanical 
contribution to the mean kinetic energy is due to the quantum uncertainty 
principle localization of the proton.

\appendix 
\section{Inequalities \label{ine}}

The mobility of a hydrogen nucleus may be expressed via the 
Kubo formula\cite{Kubo:1998} as 
\begin{equation}
\mu (\zeta )=\frac{1}{3\hbar }\int_0^\beta d\lambda \int_0^\infty dt 
\ e^{i\zeta t} \left<{\bf V}(-i\lambda)\cdot {\bf V}(t))\right> , 
\label{ine1}
\end{equation}
wherein 
\begin{equation}
{\bf V}\equiv \dot{\bf R}=\frac{i}{\hbar }\left[{\cal H},{\bf R}\right], 
\ \ \ \beta =\frac{\hbar }{k_BT}\ \ \ {\rm and}
\ \ \ {\Im}m\ \zeta >0.
\label{ine2}
\end{equation}
The mobility obeys the sum rules 
\begin{eqnarray}
\frac{2M}{\pi }\int_0^\infty {\Re }e\{{\mu}(\omega +i0^+)\} d\omega =1, 
\nonumber \\ 
\frac{2M}{\pi }\int_0^\infty \omega^2 {\Re }e\{{\mu}(\omega +i0^+)\} d\omega 
=\Omega^2 \equiv \frac{4\pi ne^2}{3M}\ , 
\nonumber \\ 
n=\left<\sum_j \delta ({\bf R}-{\bf r}_j)\right>.
\label{ine2_prime}
\end{eqnarray}
The self diffusion coefficient obeys the Einstein relations 
\begin{eqnarray}
D_{\rm self}= (k_BT) \lim_{\omega \to 0} {\Re }e\{ \mu (\omega +i0^+)\}, 
\nonumber \\
D_{\rm self}=\lim_{t\to \infty }
\left<\big({\bf R}(t)-{\bf R}(0)\big)^2\right>/(6t)\ .
\label{ine3}
\end{eqnarray}
The fluctuation-response theorem\cite{Kubo:1998} implies a rigorous 
expression for the kinetic energy of the hydrogen nucleus 
\begin{eqnarray}
\frac{1}{2}M\left<|{\bf V}|^2\right>&=& 
\frac{3}{2}\left\{\frac{2M}{\pi }
\int_0^\infty E_T(\omega ){\Re }e\{ \mu (\omega +i0^+)\} d\omega\right\}, 
\nonumber \\ 
E_T(\omega )&=&\left(n(\omega )+\frac{1}{2}\right)\hbar \omega ,
\nonumber \\ 
E_T(\omega )&=&
\left(\frac{\hbar \omega }{2}\right)
\coth\left(\frac{\hbar \omega }{2k_BT}\right),  
\label{ine4}
\end{eqnarray} 
wherein \begin{math} E_T(\omega ) \end{math} is the mean energy of 
a thermal oscillator of frequency \begin{math} \omega  \end{math}.
The inequality 
\begin{equation}
E_T(\omega )\ge k_BT 
\label{ine5}
\end{equation}
together with Eqs.(\ref{ine2_prime}) and (\ref{ine4}) imply   
\begin{eqnarray}
\frac{1}{2}M\left<|{\bf V}|^2\right>
&\equiv & \frac{3}{2}k_B\tilde{T},
\nonumber \\ 
\tilde{T} &\ge & T. 
\label{ine6}
\end{eqnarray}
For a particle obeying classical mechanics to a sufficient degree of 
accuracy, equality \begin{math} \tilde{T} = T  \end{math} in 
Eq.(\ref{ine6}) holds true. For a particle obeying quantum mechanics, 
the quantum fluctuations yield kinetic energies over and above the classical 
thermal value \begin{math} \tilde{T} > T  \end{math}.

\section*{Acknowledgments}

J. S. would like to thank the United States National Science Foundation for support under PHY-1205845.

\end{document}